# Retail Price Ripples *

Xiao Ling**
Assistant Professor of Marketing
School of Business, Central Connecticut State University
New Britain, CT 06050, USA
x.ling@ccsu.edu

Sourav Ray
Lang Chair and Professor of Marketing
Gordaon S. Lang School of Business and Economics
University of Guelph, ON N1G 2W1, Canada
s_ray@uoguelph.ca

Daniel Levy
William Gittes Chair, Department of Economics
Bar-Ilan University, Ramat-Gan 5290002, Israel
Emory University, Atlanta GA 30322, USA
ICEA, Wilfrid Laurier University, Canada
ISET at TSU, Tbilisi, Georgia
RCEA, University of Bologna, Rimini, Italy
daniel.levy@biu.ac.il

---

Working paper version

Last Revision:
December 1, 2025

---

---

**HIGHLIGHTS**

1. Small retail price changes can have big implications for retailer profitability.

2. Small price changes across the retail basket can accumulate to a big impact for consumers' out-of-pocket expenses.

3. Observed small price changes can be due to inflation or a deliberate retail tactic that leverages consumer inattention.

4. Retailers can profitably leverage consumer inattention with small price increases or conceal large price increases among many small decreases.

5. Economy-wide study documents systematic small price changes in grocery retail that cannot be attributed to inflation, pointing to a deliberate retail pricing tactic.

# Retail Price Ripples

## Abstract

Much like small ripples in a stream, which get lost in the larger waves, small changes in retail prices often fly under the radar of public perceptions, while large price changes appear as marketing moves associated with demand and competition. Unnoticed, these could increase consumers' out-of-pocket expenses. Indeed, retailers could boost their profits by making numerous small price increases or by obfuscating large price increases with numerous small price decreases, thereby bypassing the consumer's full attention and consideration, and triggering consumer fairness concerns. Yet only a handful of papers study small price changes. Extant results are often based on a single retailer, limited products, short time span, and legacy datasets dating back to the 1980s and 1990s – leaving their current practical relevance questionable. Researchers have also questioned whether the reported observations of small price changes are artifacts of measurement errors driven by data aggregation. In a series of analyses of a large dataset (almost 79 billion weekly price observations from 2006 to 2015, covering 527 products, and about 35,000 stores across 161 retailers), we find robust evidence of asymmetric pricing in the small (APIS), where small price increases outnumber small price decreases, but no such asymmetry is present in the large. We also document the reverse phenomenon (APIS-R), where small price decreases outnumber small price increases. Our results are robust to several possible measurement issues. Importantly, our findings indicate a greater current relevance and generalizability of such asymmetric pricing practices than the existing literature recognizes.

# 1.    INTRODUCTION

The prices consumers pay at retail are often at the forefront of their concerns about fair dealing.  This is certainly true if price changes are implemented in a manner that can potentially bypass the consumer's full attention and consideration.  Comprehensively establishing if any presumed standard of fairness has been breached is a complex and time-consuming process, but such pricing raises an interesting challenge for researchers.  If these pricing patterns escape consumer attention and consideration, they will not become dominant public concerns unless the research evidence suggests widespread use and stable patterns of such pricing behavior in our retail ecosystem.  So, estimating and documenting generalizable parameters and empirical evidence of pricing patterns that can be seen as bypassing the consumer's full attention and consideration is critical for informed public discourse.  The focus of this paper is to establish the evidence of exactly such pricing patterns in our grocery retail ecosystem.

Retailers often face intense pressure to raise prices, creating a critical dilemma: is it better to use a single, transparent price hike, or a series of smaller, 'under-the-radar' price adjustments?  Small price changes like these, which we call "retail price ripples," rarely make headlines, but they can potentially move baskets and margins, especially in highly competitive markets.  Yet, when retailers adjust prices by just a few cents, customers often do not react, and the changes can accumulate to a meaningful profit impact.  At the same time, repeated micro-increases without matching decreases will accumulate, which may trigger perceptions of unfairness if they appear to creep in "under the radar," especially at inflation times when consumers already feel squeezed.

Indeed, while for economists, small price changes are seen as key to the general response to inflation, deflation, etc., for retailers, small price adjustments can be a strategic tool with implications for profitability and competitive posturing. Therefore, evidence that such price ripples are endemic retail pricing practices can compound the sense of unfairness.

Yet, despite the significant business and policy implications associated with them, only a handful of papers study small price changes explicitly. Even this limited set of studies has significant limitations that raise questions about the robustness and generalizability of their empirical findings. In particular, existing evidence often relies on single retailers, narrow product ranges, short observation windows, or legacy data from the 1980s and 1990s (e.g., Carlton, 1986; Lach and Tsiddon, 2007; Ray et al., 2006; Chen et al., 2008; Levy et al., 2025). Recent work has also questioned whether observed asymmetries reflect true firm behavior or perhaps they are artifacts of data aggregation and measurement errors (Eichenbaum et al., 2014; Campbell & Eden, 2014).

This study provides the first large-scale and robust evidence on these issues and establishes a baseline for understanding how small price changes shape competitive and consumer dynamics in modern retailing. Using a large, comprehensive dataset that covers much of the US geographic territory, we re-examine the ripples (small changes) in grocery retail prices. Specifically, we document the phenomenon of asymmetric pricing in the small (APIS), where there are statistically more small price increases (penny rises) than small price decreases (penny drops), and where such asymmetry between positive and negative price changes tends to vanish for larger changes. The corresponding reverse phenomenon, where there is statistically greater number of small price decreases than small price increases, is asymmetric pricing in the small – reversed (APIS-R).

Our key research questions are: Do small price increases outnumber small price decreases in retail price spectrum? How prevalent are APIS and APIS-R in the spectrum of current retail pricing practices? Are these mere artifacts of inflation and measurement problems? What is the scale and scope of these phenomena after controlling for inflation and some key measurement concerns? We contend that these are important questions and expect the answers will help researchers in marketing and economics focus further studies in the domain.

To address these questions, we estimate APIS and APIS-R in a dataset comprising more than 79 billion weekly retail price observations over a relatively recent 10-year period (2006–2015). The data covers a broad range of 56 categories (product groups) and 527 sub-categories (product modules) sold in 35,000 stores belonging to 161 retailers in the US. We also use an independent transactions dataset to simulate the impact of price aggregation on the estimation of APIS and APIS-R.

We make three specific contributions. First, we estimate and document APIS in retail grocery, a key sector for the economy. This is important, for there are only a handful of papers (Ray et al., 2006; Chen et al., 2008; Levy et al., 2025) that document the APIS phenomenon. We find that APIS continues to be a part of the retail-pricing spectrum even after two decades of the data reported in the first study and even after major economy-wide technological changes that could reasonably be seen as affecting consumer decision making, and by implication, retail pricing practices. Second, we document stable patterns of APIS-R. Chakraborty et al. (2015) is the only other paper that reports this. This extends the asymmetric pricing literature by expanding the documented spectrum of retail pricing practices. Third, we show that the observed APIS and APIS-R findings are robust to several measurement concerns – in particular, to unit value indices (UVIs) – pointed out as a key source of noise by Eichenbaum et al. (2014) and Campbell and Eden (2014). To the best of our knowledge, our database of grocery retailers across the nation is the largest dataset to be used in studying small price changes to date. Thus, our work also lends more generalizability to the documented phenomena. Our findings are robust across several bases of aggregation – increasingly granular categorization of products (sub-category, category, and department), retailers, time, location, etc.

In the following section, we elaborate on the motivation behind this research, outlining the key theoretical perspectives, identifying the tensions arising out of conflicting results, and formalizing our key research questions. We follow in section 3 by discussing the trade-offs of using aggregate versus transaction data for our research, our approach to dealing with the challenges, and describing the new datasets used. Next, in section 4, we discuss measurement and estimation, followed in section 5 by the findings, in section 6 by discussions of the results, and by conclusions in section 7.

## 2. MOTIVATION

### 2.1 The Big Role of Small Price Changes

The general phenomenon of small price changes is important for marketing for various reasons. Marketers often consider small price changes a strategic choice with implications for profitability and competitive positioning. It is well known that customers do not always perceive price changes and may ignore price changes either rationally or irrationally. Consumers' perception of reference prices is a function of prices they see (Mayhew & Winer, 1992; Greenleaf, 1995). In such cases, a small price change could become a meaningful lever for marketers. Marketers could use price changes to modify the price spectrum to form customer reference prices. For example, small price changes may be used by retailers to strategically obfuscate their price spectrum while at the same time maintaining their (competitive) price image (Chakraborty et al., 2015).

Small price changes are important for economics as well. In fact, macroeconomists have considered small price changes in the context of a general response to inflation, deflation, etc. They are tied to the issue of price rigidity and the associated monetary policy concerns. Economists have also examined asymmetric price changes ("rockets and feathers"): prices rising faster than they fall, in contexts such as gasoline and commodities (e.g., Peltzman, 2000; Meyer & von Cramon-Taubadel, 2004). Unlike these macro-level studies of price transmission, our focus is on micro-level price adjustments in retail settings.

## 2.2    *Multiplicity of perspectives and findings*

Four main theoretical perspectives drive much of the studies of small price changes in marketing. The *price adjustment cost or "menu cost" (MC)* line of work focuses on the retailer's (in)ability to change prices on account of the associated costs (Barro, 1972; Sheshinski and Weiss, 1977, 1979; Mankiw, 1985; Levy et al., 1997, 1998; Dutta et al., 1999). The *Rational Inattention (RI)* literature (e.g., Reis, 2006a, b; Mankiw & Reis, 2002, 2011; Ball et al., 2005; Sims, 2003, 2010; Chen et al., 2008) and *just-noticeable differences (JND)* literatures (Kalyanaram & Little, 1994; Lichtenstein, Block, & Black, 1988; Gupta & Cooper, 1992; Fibich et al., 2007; Pauwels et al., 2007) focus on customer motivation and ability to process small price changes as a precursor to their consumption decisions. Studies in this literature argue that when price reductions are within certain (not necessarily symmetric) thresholds, consumers do not react to them by changing their purchase behavior. An emerging fourth stream of research argues that small price adjustments are tactical in nature and are an essential part of the marketer's *strategic dynamic pricing efforts* (Ray et al., 2006; Ray et al., 2012; Wood et al., 2013; Chakraborty et al., 2015; Levy et al., 2025).

The multiple perspectives to interpret small price changes have generated a rich set of research insights for researchers, policy makers, and marketers. However, the different perspectives used to study small price changes have conflicting predictions regarding their desirability and profitability for retailers. While the menu cost line of reasoning argues against the existence of any small price changes – increases or decreases, the RI and JND perspectives generally predict against the existence of small price decreases as part of a retailer's pricing practice since these will not be profitable for the retailer. This prediction, of course, stands in contrast to the strategic obfuscation line of reasoning of Chakraborty et al. (2015) who argue for the existence of numerous small price decreases, counter to the proposition of Chen et al. (2008) and Levy et al. (2025). Therefore, on the face of it, consensus weighs against observing systematic evidence of any type of small price changes, whether positive or negative. However, that is not the case. Existing empirical studies document the presence of frequent small price changes, both increases and decreases, in micro-level transactions price data (e.g., Carlton 1986; Lach & Tsiddon, 2007; Ray et al., 2006; Ray et al., 2012; Wood et al., 2013; Chakraborty et al., 2015). These conflicts certainly lead to questions about the robustness of previous findings. To this end, existing research has important limitations.

### 2.3    Data Limitations

Much of the limitations referred to above have to do with data. Firstly, existing results are based largely on limited samples – in terms of both the number of stores and the number of products sampled. Further, much of the direct evidence is based on relatively old data collected in the late 1980s and early 1990s, before the advent of the internet and e-commerce (e.g., Carlton, 1986; Lach and Tsiddon, 2007; Ray et al., 2006; Chen et al., 2008; Levy et al., 2025). Overall, this limited nature of the data constrains the inferences we can draw about the generalizability and contemporary relevance of the findings. To add to this challenge, Eichenbaum et al. (2014) and Campbell & Eden (2014) have raised concerns about the measurement of small price changes in aggregated data. They contend that much of our observations of small price changes might be artifacts of aggregation and other data handling practices when data collectors convert transaction-level price data into aggregated scanner price data. Therefore, re-examining small price changes for generalizability as well as the impact of measurement artifacts is in order.

### 2.4    Relevance of APIS and APIS-R in the retail pricing spectrum

A core element of the domain of small price changes is the phenomenon of asymmetric pricing in the small (APIS and APIS-R). These are important for the marketers inasmuch as they are strategically deployed dynamic pricing tactics by the retailer. Nevertheless, only a few papers directly study the phenomena, and even the findings of this small set can be contradictory. In fact, to the best of our knowledge, only Chen et al. (2008), Chakraborty et al. (2015), and Levy et al. (2025) specifically study APIS.

While Chen et al. (2008) and Levy et al. (2025) report robust evidence of APIS and none for APIS-R, Chakraborty et al. (2015) find robust evidence of APIS-R as well. These discrepancies may arise partly from the limited scope of their datasets, with Chen et al. and Levy et al. relying on pricing data from a single U.S. retailer, and Chakraborty et al. drawing on three U.K. grocery chains[1]. This paucity of studies in the domain and heterogeneity in results further amplify the limitations identified earlier. Therefore, how generalizable APIS and APIS-R are, in terms of their prevalence in the economy, and their current relevance as part of the retailer's spectrum of pricing practices, are important research questions. Specifically, what is the scale and scope of these phenomena after controlling for some of the measurement concerns raised in the literature? We contend that these concerns are non-trivial for the literature, and we expect that a more definitive answer to these questions will help researchers in marketing advance studies in this domain further.

---

[1] In contrast, the dataset used in this study comprises pricing information from 161 of the largest U.S. retail chains spanning more than 79 billion weekly price observations, offering substantially greater breadth for evaluating generalizability.

### 2.5 Research Questions

Our review of the literature reveals significant tensions regarding the existence, prevalence, and nature of asymmetric pricing in the small. First, theoretical perspectives conflict: menu cost models predict no small changes, while RI/JND models support APIS but not APIS-R. Second, empirical findings are mixed: Chen et al. (2008) and Levy et al. (2025) find APIS but not APIS-R, while Chakravarty et al. (2015) find robust evidence for both. Finally, recent methodological critiques suggest these findings may be mere "artifacts" of data aggregation (e.g., UVIs) or inflation. To adjudicate these conflicts and provide a large-sample, contemporary test, we formalize our study around three key research questions:

RQ1: To what extent do small price increases systematically outnumber small price decreases (APIS), and, in parallel, do small price decreases systematically outnumber small price increases (APIS-R), across a broad spectrum of U.S. grocery retailers?

RQ2: Are these observed APIS and APIS-R patterns robust phenomena, or are they primarily artifacts of inflation, disappearing during periods of low inflation or deflation?

RQ3: Do the APIS and APIS-R patterns persist after controlling for potential measurement artifacts, such as 1¢ rounding, bundle pricing, and data aggregation (UVIs), that have been suggested as sources of spurious small price changes?

### 3.    DATA

*3.1    Aggregate versus Transactions data for our study*

Researchers studying retail pricing using secondary data generally have access to two types of datasets.  True *transaction datasets* are generated at the retail point of sale.  *Aggregate datasets* are compiled from multiple transactions – e.g., prices of all transactions in a day (week) can be aggregated to create a daily (weekly) average price for the product.  While there are a limited number of small price change studies with *transaction data* (cf. Ray et al., 2012; Wood et al., 2013; Chakraborty et al., 2015), most utilize *aggregated price data* (cf. Ray et al., 2006; Lach & Tsiddon, 2007; Chen et al., 2008; Levy et al., 2020; Levy et al., 2025).

Not surprisingly, both data types have their advantages and disadvantages for studying our research problem.  Aggregated data are more widely available and typically cover broader product ranges, more retailers, and longer time periods, making them well-suited for assessing prevalence and generalizability (Besanko et al., 2003; Nakamura & Steinsson, 2013).  However, aggregation may introduce measurement noise, especially when prices are averaged across time (Eichenbaum et al., 2014; Campbell & Eden, 2014; Cavallo, 2018).  Transaction-level datasets, while more precise, are less commonly available and tend to cover fewer stores, product categories, and time periods, limiting their external validity.  Given our research goals, which prioritize generalizability while remaining attentive to measurement concerns, we adopt a multi-pronged empirical strategy that combines both data types.

*3.1    Data Description*

We use the NielsenIQ Retail Scanner Dataset (hereinafter referred to as NRS dataset), an aggregated dataset, as our main data for the analyses, supplemented by a separate transaction price dataset for an additional analysis.  The NRS data is a panel dataset of total sales (quantities and prices) at the UPC (barcode) level for around 35,000 geographically dispersed stores belonging to more than 160 retail chains (these numbers vary by year) across all US markets[2]. The data consists of *weekly* pricing, volume, and store-merchandising information aggregated from transactions recorded by the stores' point-of-sale systems from 2006 to 2015. The strength of this data is evident: it provides a geographically economy-wide context in a contemporary setting, enabling the best generalizability possible. Figure 1 shows the geographic coverage of our data.

The NRS data organizes the product hierarchy into ten product departments, which are then further organized into 125 product groups, followed by product modules and the individual SKUs or UPCs. The product hierarchy of our sample is organized into 9 randomly chosen product departments, which are then further organized into 56 randomly chosen product groups consisting of 527 product modules (sub-categories), comprising 4,311,648 UPCs. For ease of referencing, we will hereinafter refer to the groups as categories and the modules as sub-categories.

Figure 2 shows the hierarchy of the categorization. Alcohol and tobacco products are excluded because those products are heavily regulated in the US. The selected 56 categories with 527 sub-categories cover most of the product categories studied in previous research, including most of the 27 categories covered by Chen et al. (2008) and Levy et al. (2025). Our sample comprises 161 retailers, belonging to 91 parent companies. This represents a majority of retailers recorded in the full database. The data sample, in total, contains more than 79 billion weekly price observations. The summary of category-level observations is reported in Table 1.

---

[2] The full dataset covers more than 50% of the total sales volume of US grocery and drug stores and more than 30% of all US mass merchandiser sales volume.

As a robustness check of the claims against data aggregation, we also analyze an independent, smaller transaction price dataset, which consists of two stores in the North-West Milan region of Italy. We discuss this later in the analysis.

(FIGURE 1 and 2 ABOUT HERE)

### 4. MEASUREMENT AND ESTIMATIONS

*4.1 Asymmetric Pricing in-the-Small (APIS and APIS-R)*

Following Chen et al. (2008), we analyze the frequency of both positive and negative price changes at different aggregation levels (e.g., category, retailer etc.). We compute weekly unit prices (in cents) and track week-to-week differences for each product in the same store, capturing both upward and downward price adjustments. The positive and negative price change frequencies are computed for each possible size of price change in cents: 1¢, 2¢, 3¢, etc., up to 1,000¢. For example, we calculate the frequency of 5¢ price increases (and decreases) during the whole year (or the 10-year period) for all the items in a certain product module.

In Figure 3, we plot the cross-category aggregated frequency of positive and negative price changes throughout the 10-year sample period in the NRS data. This reveals the pattern of APIS – a greater number of small price increases than decreases. This asymmetry exists for a range of up to about 10¢ of price-change, which is in line with the findings of Chen et al (2008) and Levy et al (2025). Beyond that range, the asymmetry disappears as the two lines start crisscrossing each other and the differences between positive and negative price change frequencies gradually converge to zero. Note also that the most frequent price change magnitudes are multiples of 10¢ – 10¢, 20¢, 30¢, 50¢, $1, $1.5, etc., which matches what Levy et al. (2011) find in their study of the 9-ending pricing phenomenon.

(FIGURE 3 ABOUT HERE)

Following Chen et al. (2008), we use "asymmetry threshold" as a measurement of the magnitude of asymmetry. We define the asymmetry threshold as the magnitude of price change below which asymmetric pricing is statistically supported. Our primary threshold measurement method is to aggregate price change frequencies at different price-change magnitude and to find the first point where the pricing asymmetry does not hold, i.e., the first point (from 0 cent price change to 1000 cents price change) where no statistical difference between the positive and negative changes is observed. We use the Z-test to measure the statistical significance (at a 5% significance level) of the probability that there is no asymmetry at each price change point. Hence, the first point (from 0 cent price change to hundreds of cents price change at an interval of 1 cent), where the null hypothesis of symmetric price change cannot be rejected, or the first point where the direction of asymmetry changes, is defined as the asymmetric price-change threshold. A non-zero price change asymmetry threshold can be found as evidence of either APIS or APIS-R, depending on the direction of asymmetry within the threshold point.

## 5.    ANALYSES AND RESULTS

In this section, we present the empirical results. We first measure the APIS and APIS-R thresholds at different aggregation levels using the NRS dataset to establish the prevalence of APIS and APIS-R (RQ1). We then test the robustness of this finding by controlling for the effects of inflation (RQ2). Finally, we conduct a series of tests to ensure our findings are not driven by the measurement errors, including simulations with transaction-level data (RQ3).[3]

*5.1    Prevalence of APIS and APIS-R (RQ1)*

---

[3] Several additional robustness checks, including the analyses of lagged price adjustments, are detailed in the Web Appendix.

To answer RQ1, we first analyzed the full sample of the NRS data. Across all aggregation levels, both APIS and APIS-R patterns appear consistently, though the magnitude of asymmetry varies by category. APIS is more prominent as the aggregation level goes up: the average threshold and the proportion of APIS increases significantly at the category and department levels compared with at the sub-category level. Detailed thresholds by category and department are reported in Tables 1 and 2, with a module-level summary in the Web Appendix.

Overall, APIS patterns dominate at all aggregation levels. At the sub-category level, 247 out of 527 (46.9%) sub-categories exhibit APIS thresholds, whereas 164 sub-categories (31.1%) exhibit APIS-R; the rest 116 (22%) have no asymmetry (i.e., the asymmetry threshold is 0). For most sub-categories with APIS, the asymmetry thresholds fall in the range of positive 2¢ to 30¢. The average thresholds are about 18.1¢ and 7.4¢ for APIS and APIS-R, respectively. The overall average asymmetry threshold is about 6.2¢. Figure 4 plots the distribution of asymmetry thresholds at the sub-category level.

At broader aggregation levels, product categories and departments, the asymmetry becomes more pronounced. At the product category level, 31 categories (55.4%) have an APIS threshold, and 22 categories (39.3%) have APIS-R thresholds (see Table 1). The average APIS threshold is 27¢ at the category level, 8.9¢ higher than at the sub-category level. The average APIS-R threshold is 4.7¢ at the category level, 2.7¢ higher than at the sub-category level. The higher APIS thresholds are most pronounced at the department level, where the average APIS threshold is 34.2¢ (see Table 2).

At the retailer level, 83% of chains display some form of small-change asymmetry, predominantly APIS (93 retailers) and, to a lesser extent, APIS-R (40 retailers). 59 retailers (36.6%) exhibit APIS thresholds of 10¢ or higher. In contrast, most APIS-R thresholds (22 out of the 40) are smaller than 5¢. Figure 5 plots the distribution of asymmetry thresholds using the Retailer Code to identify the retail chains.

(TABLE 1, FIGURE 4 AND FIGURE 5 ABOUT HERE)

*5.2    The Role of Inflation (RQ2)*

To address RQ2, we test whether the observed asymmetries are merely an artifact of inflation, as suggested by Ball and Mankiw (1994). We control for inflation by applying the method of Chen et al. (2008), creating three sub-samples based on monthly PPI inflation rates: low-inflation, deflation, and inflation periods.

The low-inflation period sample only includes observations during which the monthly PPI inflation rate was positive but did not exceed 0.1% (only two months are identified as low-inflation period during the 2006–2015 period: August 2010 and June 2013). The low-inflation sub-sample is more conservative, in which only months with a non-positive inflation rate are included (the PPI inflation rate used is not seasonally adjusted). This sample covers 49 months with an average monthly inflation rate of about −1.06%. The third sample, which covers 47 months with an average monthly inflation rate of about 1.3%, is defined as the inflation period sample, where only months with a PPI inflation rate higher than 0.5% are kept. The analysis results at the product department level are summarized in Table 2. See Table A3 in the Web Appendix for results at the product category level.

The results demonstrate that the asymmetries are not driven solely by inflation. In the deflation-period sample, APIS is still present in 218 out of 527 sub-categories (41.3%), and APIS-R becomes slightly more prominent. Conversely, during high-inflation months, 26.9% of sub-categories still exhibit APIS-R. At the category level, the average threshold decreases during both the low inflation (from 13.1¢ to 3¢) and the deflation (to 6.9¢) periods. Inflation clearly influences, but does not determine, the formation of APIS. Even during deflationary months, both APIS and APIS-R persist across a large share of categories. Further, APIS categories always tend to have larger thresholds than APIS-R, even during deflationary periods. We thus conclude that the asymmetry "in the small" is a robust finding and not simply an artifact of inflation, providing a clear answer to RQ2.

(TABLE 2 ABOUT HERE)

### 5.3     Robustness to Measurement Artifacts (RQ3)

Our third research question (RQ3) addresses the critical challenge that APIS may be a spurious artifact of data measurement during the aggregation of transaction-level price data. Recent literature suggests these artifacts arise from three main sources: (1) rounding errors (especially 1¢ changes), (2) bundle pricing and unit value indices (UVIs) creating fractional prices, and (3) time aggregation of transaction data. We address these concerns in two complementary ways: First, we re-estimate asymmetries on a filtered subsample of the main dataset to remove potential artifacts (see results in Table 1 and Table 2). Second, we validate the results using an independent transaction-level dataset that allows direct comparison before and after aggregation.

As our primary test for RQ3, we created a conservative subsample of the NRS data designed to filter out these artifacts. To create a conservative 'clean' sample, we removed bundle-priced items (price multiplier > 1), fractional prices caused by UVIs, and implausibly small or large changes (< 0.1% or > 120%), following Alvarez et al. (2014)[4].

---

[4] Alvarez et al. (2014) exclude extreme price changes as well as smaller than 1¢ price changes, in their analyses of two scanner data sets (Dominick's and IRI scanner dataset). They find that the resulting distribution of price changes matches the distribution found in datasets which are immune to this kind of errors. We take their method further by not allowing for any fractional price changes.

The results from this "clean" subsample are conclusive: systematic pricing asymmetry "in the small" persists. We find almost the same number of APIS (31) and APIS-R (23) thresholds at the category level compared to the full sample (31 and 22, respectively). The number of the categories with no asymmetry drops slightly from 3 to 2, which means that more than 96% of the categories still exhibit asymmetry. At the department level, we still observe 6 APIS and 3 APIS-R out of the 9 departments, with an average APIS threshold of 25.5¢ and an average APIS-R threshold of 5¢. This implies that spurious prices may influence the measurement of asymmetry thresholds to a small extent, but do not account for the vast majority of the pricing asymmetry we observe. This provides strong evidence that the asymmetry "in the small" regularity is a robust phenomenon and not a product of these measurement errors.

### 5.4 Transaction-Level Validation (External Robustness)

As our secondary test for RQ3, we analyze a separate transaction-level dataset from two grocery stores operated by a major retailer in Northwest Milan. Both stores belong to one of the largest grocery chains in the country. Store A is a High-Low (HILO) type store, and Store B is an Every Day Low Price (EDLP) store. The data includes sales receipts generated at the point of sale over about one year starting July 2007 – around 2.6 million price observations for Store A and 3.4 million for Store B,[5] with no weekly or modular aggregation. Applying the same threshold estimation procedure, we find evidence of both APIS and APIS-R, consistent with our earlier findings, confirming that the patterns are not artifacts of aggregation.

We further simulate aggregation effects to test whether aggregating transactions at the weekly level alters the estimated asymmetry. We convert the transaction price observations in our two-store data into weekly weighted average prices, the same way the NRS dataset was handled. After conversion, we observe that aggregation does not systematically increase the frequency of small price changes,[6] contrary to concerns raised in prior research (Eichenbaum et al., 2014). See Figure 6 for a comparison of price change frequency distribution between transaction level and aggregated level.

---

[5] Data of Store A covers the period from July 2nd, 2007, to October 27, 2008, a total of 68 weeks. Data of Store B covers the period from July 16, 2007, to September 8th, 2008, a total of 59 weeks.

(FIGURE 6 ABOUT HERE)

Whether estimated from daily transaction prices or weekly aggregates, the asymmetry patterns persist across stores and aggregation levels. Differences in asymmetry thresholds before and after aggregation are minimal and not statistically significant based on two-sample *t*-tests. In some cases, aggregation slightly underestimates the magnitude of asymmetry. The results are summarized in the Web Appendix, Table A4 and Table A5. Taken together, these results confirm that APIS and APIS-R are not artifacts of aggregation or measurement noise but robust features of retail price setting.

Notably, the transaction-level data from Italy were collected during a period of moderate inflation, with the annual inflation rate rising from approximately 2.2% in July 2007 to 3.8% in June 2008. Despite this inflationary environment, we observe both APIS and APIS-R patterns across stores and product categories. This finding, again, suggests that inflation alone does not fully explain the observed asymmetries. Instead, the persistence of both APIS and APIS-R even during an inflationary period supports the interpretation that asymmetric pricing in the small reflects intrinsic retailer pricing behaviors rather than being purely macroeconomic artifacts.

*5.4    Additional Robustness Tests*

We conducted several additional robustness tests (summarized below; full details in the Web Appendix). These tests confirm that: Excluding all 1¢ price changes does not eliminate the asymmetry; in fact, it significantly increases the proportion of APIS thresholds, suggesting 1¢ rounding may have been underestimating the effect. (See Web Appendix, Analysis A1). The results are also robust to alternative measures of inflation (CPI instead of PPI), the use of lagged price adjustments, and the first-year sample versus the last-year sample. (See Web Appendix, Table A2, A3, and A6).

---

[6] We did the comparison using paired-sample T-tests.

## 6.    DISCUSSION

Our study was primarily motivated by a desire to look for robust and stable evidence of asymmetric pricing in the small (APIS and APIS-R) – pricing patterns that can be seen as bypassing the consumer's full attention and consideration.  Three research questions formed the backdrop of our study.  Our results provide clear answers: In response to RQ1, we find contemporary, large-scale evidence for both APIS and APIS-R, suggesting that both the RI/JND and strategic obfuscation perspectives are at play in the market.  In response to RQ2 and RQ3, we demonstrate that these phenomena are not artifacts of inflation or data measurement.  Taken together, we believe our findings move the dial significantly to establishing the generalizability of the phenomena of asymmetric pricing in the small.  In this section, we discuss the theoretical, managerial, and policy implications of these findings.

### 6.1  Theoretical Interpretation

This study contributes to the literature by reconciling several long-standing tensions across major perspectives on small price changes.  Four perspectives dominate studies of small price changes: the "menu cost" line of work, the rational inattention (RI) approach, the just-noticeable-difference (JND) literature, and the strategic intent argument.  These theories, however, do not always converge on the basic question of whether small price changes will even exist.  For example, the RI and JND theories offer a rationale for the existence of small price increases but do not offer any predictions for small price decreases. The menu costs theory generally predicts against small price changes per se. whether positive or negative. Together, they seem to rule out small price changes in general, certainly small price decreases (e.g., Dutta 1999, Ray et al. 2012, Levy et al. 2020, Gupta and Cooper 1992). The strategic obfuscation line of reasoning of Chakraborty et al. (2015), on the other hand, predicts small price decreases as a tactical behavior of profit-seeking sellers.

Our results provide robust evidence of asymmetric small price adjustments in both directions. The prevalence of APIS is consistent with RI- and JND-based explanations, suggesting that consumers' limited attention to small price changes allow retailers to raise prices incrementally without substantial behavioral response. Conversely, the presence of APIS-R patterns aligns with the strategic obfuscation logic, in which small price decreases may reinforce a retailer's value image or divert attention from larger price increases.

While our analyses cannot empirically isolate these mechanisms, the coexistence of both patterns indicates that behavioral and strategic motives likely operate side by side in the marketplace. These findings, therefore, suggest complementarity rather than conflict among existing theoretical perspectives. We do not rule out the role of menu costs or other frictions; rather, our results point to an equilibrium where bounded rationality, managerial intent, and adjustment costs jointly shape pricing behavior. Specifically, the results reframe otherwise small price changes (the ripples in retail prices) as something more than just random noise in price adjustments. In doing so, our results point out the importance of integrating attention-based and strategic perspectives into a unified understanding of small price changes.

*6.2 Managerial Implications*

For managers, the results underscore that cents matter. It is especially important in today's inflationary environments, where retailers must constantly adjust prices without provoking consumer backlash. The asymmetries we document are concentrated in small price movements - typically below 30¢ for APIS and below 10¢ for APIS-R. Although individually small, these price changes can accumulate into significant margin effects. In our NRS data, for instance, small price changes of 17¢ or less account for over 10% of observed price movements, representing hundreds of millions of dollars in cumulative price adjustments (see Appendix for calculations).

From a marketing strategy perspective, small, incremental price changes represent an under-appreciated lever for profit optimization and strategic posturing. Upward micro-adjustments may allow firms to boost margins without consumer backlash, while downward moves may serve as obfuscations to enhance retailers' "value" images.

Yet the strategic use of small price changes is not without risk: practices that exploit consumer inattention could trigger perceptions of unfairness, if cumulative effects become salient. Retailers should therefore calibrate small changes not only for economic impact but also for perceptual thresholds and timing. Dynamic pricing systems and category managers could use these findings to identify "safe threshold" of adjustment by product type or price point, ranges where consumer detection and backlash are minimal.

### 6.3 Policy Implications and Future Research

From a policy perspective, asymmetric small price changes (ripples) raise subtle but important welfare questions. At the core, these price ripples are important precisely because they likely bypass full consumer attention and consideration. While rational inattention models suggest that consumers willingly trade off cognitive effort for minor losses in purchasing power, strategic obfuscation theories raise concerns about welfare reductions when consumers are systematically distracted by small price decreases. As such, the cumulative effects of asymmetric pricing in the small warrant closer examination from a consumer welfare and regulatory perspective.

Our findings also suggest that standard inflation monitoring may understate the behavioral impact of small, asymmetric price moves. Regulators and policymakers should consider how such micro-adjustments accumulate over time, especially in categories central to household budgets.

For researchers, these results open several paths. Future studies could examine how category characteristics, consumer demographics, or digital retailing environments moderate asymmetry thresholds. Experimental and field-based work could probe how consumers perceive and respond to clustered small changes. More broadly, the coexistence of APIS and APIS-R invites integration between attention-based and strategy-based theories of pricing - a bridge that has rarely been built empirically.

## 7. CONCLUSIONS

This paper set out to answer three key research questions regarding the prevalence and robustness of asymmetric pricing in the small – pricing practices that can be seen as potentially bypassing full consumer attention and consideration. Using a large, contemporary dataset with over 79 billion price observations, we find robust, broad evidence for both asymmetric pricing in the small (APIS) and its reverse (APIS-R). These patterns persist across categories, retailers, and time periods, confirming that micro-level pricing asymmetries are stable components of the modern, post-ICT retail landscape.

Our findings portray small price asymmetries as stable, measurable patterns in retail pricing that derive from managerial intent and consumer cognition. Recognizing that the "small" is anything but trivial, our results highlight the penny ripples as a meaningful unit of analyses in the modern marketplace. Their managerial relevance exists in the profitability potential of minor, psychologically calibrated adjustments, which in turn illuminate the cognitive and strategic underpinnings of price dynamics in the market. Managers could treat micro-pricing as a deliberate tool - using small, well-timed changes where consumer attention is lowest and fairness risks are managed.

Despite our efforts, our work has certain limitations. We consider these as characteristic of research domains that are in their early stages, as is the study of asymmetric pricing in the small. One limitation is in terms of the data itself. Certainly, our estimations of asymmetry thresholds would be crisper if we had access to large-scale point-of-sale transaction prices. The computing challenges of dealing with such data would have been considered onerous even a decade back. However, continuing advances in computing and associated machine learning algorithms make working with such data more feasible now.

The bottleneck continues to be access to such transaction data, per se. An associated challenge is refining the measurement of the thresholds. While we control for several sources of noise, we call for more research, possibly using different data, modeling, and experimental approaches to address this. Future research should also develop a deeper understanding of these ripples in the price spectrum – the sources and drivers of variation in asymmetry thresholds.

# References

Alvarez, F., Bihan, H. L., & Lippi, F. (2014). *Small and large price changes and the propagation of monetary shocks* (No. w20155). National Bureau of Economic Research.

Andreyeva, T., Long, M. W., & Brownell, K. D. (2010). The impact of food prices on consumption: A systematic review of research on the price elasticity of demand for food. *American Journal of Public Health,* 100(2), 216–222.

Ball, L., & Mankiw, N. G. (1994). Asymmetric price adjustment and economic fluctuations. *The Economic Journal*, *104*(423), 247–261.

Ball, L., Mankiw, N. G., & Reis, R. (2005). Monetary policy for inattentive economies. *Journal of Monetary Economics*, *52*(4), 703–725.

Barro, Robert J. (1972). A Theory of Monopolistic Price Adjustment. *Review of Economic Studies*, *39 (1)*, 17–26.

Bell, D. R., & Lattin, J. M. (1998). Shopping Behavior and Consumer Preference for Store Price Format: Why "Large Basket" Shoppers Prefer EDLP. *Marketing Science*, 17(1), 66–88.

Bergen, M., Levy, D., Ray, S., Rubin, P. H., & Zeliger, B. (2008). When Little Things Mean a Lot: On the Inefficiency of Item-Pricing Laws. *The Journal of Law and Economics*, 51(2), 209–250.

Besanko, D., Dubé, J.-P., & Gupta, S. (2003). Competitive Price Discrimination Strategies in a Vertical Channel Using Aggregate Retail Data. *Management Science*, 49(9), 1121–1138.

Bitta, A. J. D., & Monroe, K. B. (1981). A multivariate analysis of the perception of value from retail price advertisements. ACR North American Advances.

Campbell, J. R., & Eden, B. (2014). Rigid prices: Evidence from us scanner data. *International Economic Review*, *55*(2), 423–442.

Carlton, D. (1986). The Rigidity of Prices. *American Economic Review*, *76*(4), 637–58.

Cavallo, A. (2018). Scraped data and sticky prices. *Review of Economics and Statistics*, *100*(1), 105–119.

Cavallo, A., & Rigobon, R. (2016). The Billion Prices Project: Using Online Prices for Measurement and Research. *Journal of Economic Perspectives*, 30(2), 151–178.

Chakraborty, R., Dobson, P. W., Seaton, J. S., & Waterson, M. (2015). Pricing in inflationary times: The penny drops. *Journal of Monetary Economics*, *76*, 71–86.

Chen, H. A., Levy, D., Ray, S., & Bergen, M. (2008). Asymmetric price adjustment in the small. *Journal of Monetary Economics*, 55(4), 728–737.

Dutta, S., Bergen, M., Levy, D., & Venable, R. (1999). Menu costs, posted prices, and multiproduct retailers. *Journal of Money, Credit, and Banking*, 683–703.

Eichenbaum, M., Jaimovich, N., Rebelo, S., & Smith, J. (2014). How frequent are small price changes? *American Economic Journal: Macroeconomics*, *6*(2), 137–55.

Fassnacht, M., & El Husseini, S. (2013). EDLP versus Hi–Lo pricing strategies in retailing—A state of the art article. *Journal of Business Economics*, 83(3), 259–289.

Fibich, G., Gavious, A., & Lowengart, O. (2007). Optimal price promotion in the presence of asymmetric reference-price effects. *Managerial and Decision Economics*, *28*(6), 569–577.

Greenleaf, E. A. (1995). The impact of reference price effects on the profitability of price promotions. *Marketing science*, *14*(1), 82–104.

Gordon, B. R., Goldfarb, A., & Li, Y. (2013). Does price elasticity vary with economic growth? A cross-category analysis. *Journal of Marketing Research*, 50(1), 4–23.

Gorodnichenko, Y., & Talavera, O. (2017). Price setting in online markets: Basic facts, international comparisons, and cross-border integration. *American Economic Review*, *107*(1), 249–82.

Gupta, S., & Cooper, L. G. (1992). The discounting of discounts and promotion thresholds. *Journal of Consumer Research*, *19*(3), 401–411.

Kahneman, Daniel, Jack L. Knetsch, and Richard H. Thaler (1986). Fairness as a Constraint on Profit Seeking: Entitlements in the Market, *American Economic Review*, 76 (September), 728–41.

Kalwani, M. U., & Yim, C. K. (1992). Consumer price and promotion expectations: An experimental study. *Journal of Marketing Research*, *29*(1), 90–100.

Kalyanaram, G., & Little, J. D. (1994). An empirical analysis of latitude of price acceptance in consumer package goods. *Journal of Consumer Research*, *21*(3), 408–418.

Lach, S., & Tsiddon, D. (2007). Small price changes and menu costs. *Managerial and Decision Economics*, *28*(7), 649–656.

Lal R, Rao R (1997) Supermarket competition: the case of every day low pricing. *Marketing Science* 16(1):60–80

Lattin JM, Ortmeyer G (1991). A theoretical rationale for everyday low pricing by grocery retailers. Research paper 1144, Stanford University, Stanford

Lawless, H. T., & Heymann, H. (1999). Measurement of sensory thresholds. In *Sensory Evaluation of Food* (pp. 173–207). Springer, Boston, MA.

Lee, D., Kauffman, R. J., & Bergen, M. E. (2009). Image effects and rational inattention in Internet-based selling. *International Journal of Electronic Commerce*, *13*(4), 127–166.

Levy, D, Lee, D., Chen, H.A., Kauffman, R., & Bergen, M. (2011). Price points and price rigidity. *Review of Economics and Statistics* 93 (4), 1417–1431.

Levy, D., Bergen, M., Dutta, S., & Venable, R. (1997). The magnitude of menu costs: direct evidence from large US supermarket chains. *The Quarterly Journal of Economics*, *112*(3), 791–824.

Levy, D., Chen, H., Ray, S., Charette, E., Ling, X., Zhao, W., Bergen, M., & Snir, A. (2025). Asymmetric price adjustment over the business cycle. *Economics Letters* 254, 112450.

Levy, D., Dutta, S., Bergen, M., & Venable, R. (1998). Price adjustment at multiproduct retailers. *Managerial and Decision Economics*, *19*(2), 81–120.

Levy, D., Snir, A., Gotler, A., & Chen, H. (Allan). (2020). Not all price endings are created equal: Price points and asymmetric price rigidity. *Journal of Monetary Economics* 110, 33-49.

Lichtenstein, D. R., Bloch, P. H., & Black, W. C. (1988). Correlates of price acceptability. *Journal of Consumer Research*, *15*(2), 243–252.

Long, Wen, Maria Luengo-Prado, and Bent Sørensen (2015). Heterogenous Consumer Shopping Behavior: Evidence from Retail Scanner data. Manuscript, University of Houston.

Malc, D., Mumel, D., & Pisnik, A. (2016). Exploring price fairness perceptions and their influence on consumer behavior. *Journal of Business Research*, 69(9), 3693–3697.

Mankiw, N. G. (1985). Small menu costs and large business cycles: A macroeconomic model of monopoly. *The Quarterly Journal of Economics*, *100*(2), 529–537.

Mankiw, N. G., & Reis, R. (2002). Sticky information versus sticky prices: a proposal to replace the New Keynesian Phillips curve. *The Quarterly Journal of Economics*, *117*(4), 1295–1328.

Mayhew, G. E., & Winer, R. S. (1992). An empirical analysis of internal and external reference prices using scanner data. *Journal of consumer Research*, *19*(1), 62–70.

Meyer, J., & von Cramon‐Taubadel, S. (2004). Asymmetric price transmission: a survey. *Journal of agricultural economics*, 55(3), 581‐611.

Monroe, K. B. (1973). Buyers' subjective perceptions of price. *Journal of Marketing Research*, *10*(1), 70–80.

Nakamura, E., & Steinsson, J. (2013). Price Rigidity: Microeconomic Evidence and Macroeconomic Implications. *Annual Review of Economics*, 5(1), 133–163.

Pauwels, K., Srinivasan, S., & Franses, P. H. (2007). When do price thresholds matter in retail categories?. *Marketing Science*, 26(1), 83–100.

Ray, S., Chen, H. (Allan), Bergen, M. E., & Levy, D. (2006). Asymmetric Wholesale Pricing: Theory and Evidence. *Marketing Science*, *25*(2), 131–154.

Ray, S., Wood, C. A., & Messinger, P. R. (2012). Multicomponent systems pricing: Rational inattention and downward rigidities. *Journal of Marketing*, 76(5), 1–17.

Reis, R. (2006a). Inattentive consumers. *Journal of Monetary Economics*, *53*(8), 1761–1800.

Reis, R. (2006b). Inattentive producers. *The Review of Economic Studies*, *73*(3), 793–821.

Sheshinski, E., & Weiss, Y. (1977). Inflation and costs of price adjustment. *The Review of Economic Studies*, *44*(2), 287–303.

Sheshinski, E., & Weiss, Y. (1979). Demand for fixed factors, inflation and adjustment costs. *The Review of Economic Studies*, *46*(1), 31–45.

Sims, C. A. (2003). Implications of rational inattention. *Journal of Monetary Economics*, *50*(3), 665–690.

Sims, C. A. (2006). Rational inattention: Beyond the linear-quadratic case. *American Economic Review*, *96*(2), 158–163.

Sims, C. A. (2010). Rational inattention and monetary economics. In *Handbook of Monetary Economics* (Vol. 3, pp. 155–181). Elsevier.

Wood, C. A., Ray, S., & Messinger, P. (2013). Leaving the tier: An examination of asymmetry in pricing patterns in online high-tech shops. *Lecture Notes in Business Information Processing*, *155 LNBIP*, 63–73.

Figure 1. Geographic Coverage of the NRS Data Sample

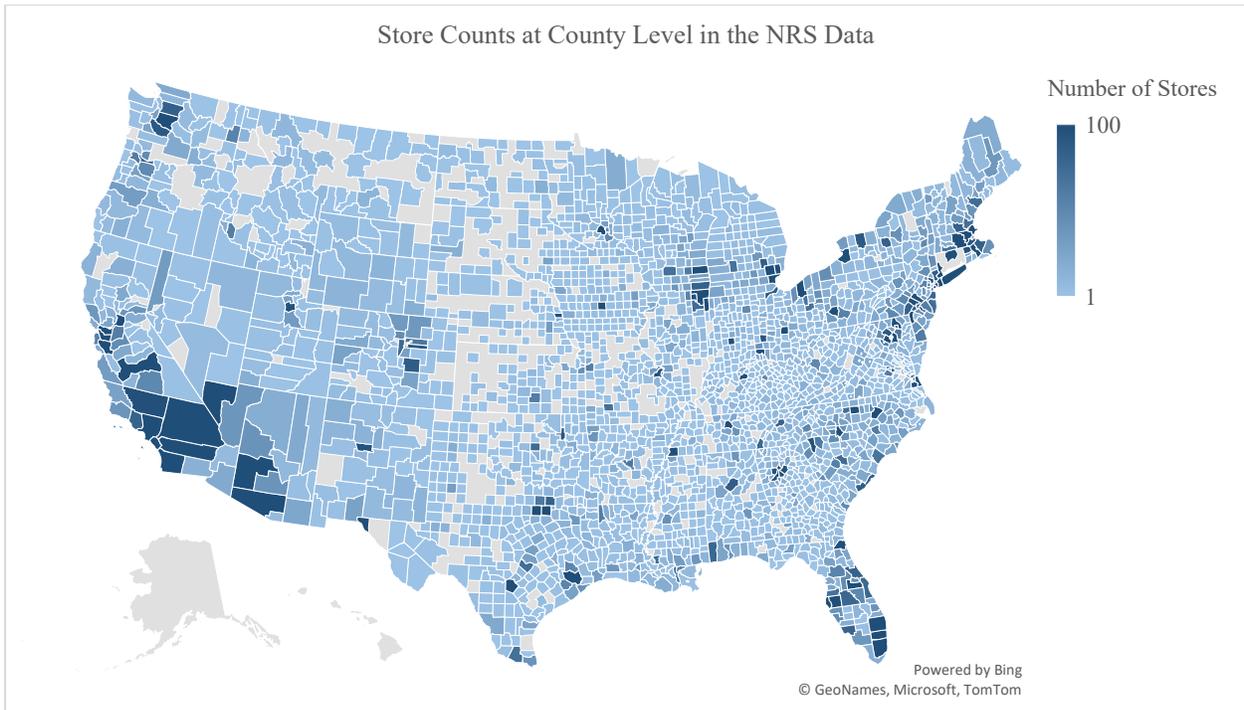

Note: The color scale is capped at 100 stores. All counties with 100 or more stores, including those with more than 700 stores, appear in the same darkest shade.

Figure 2. Hierarchy of the Product Categorization in the NRS Data

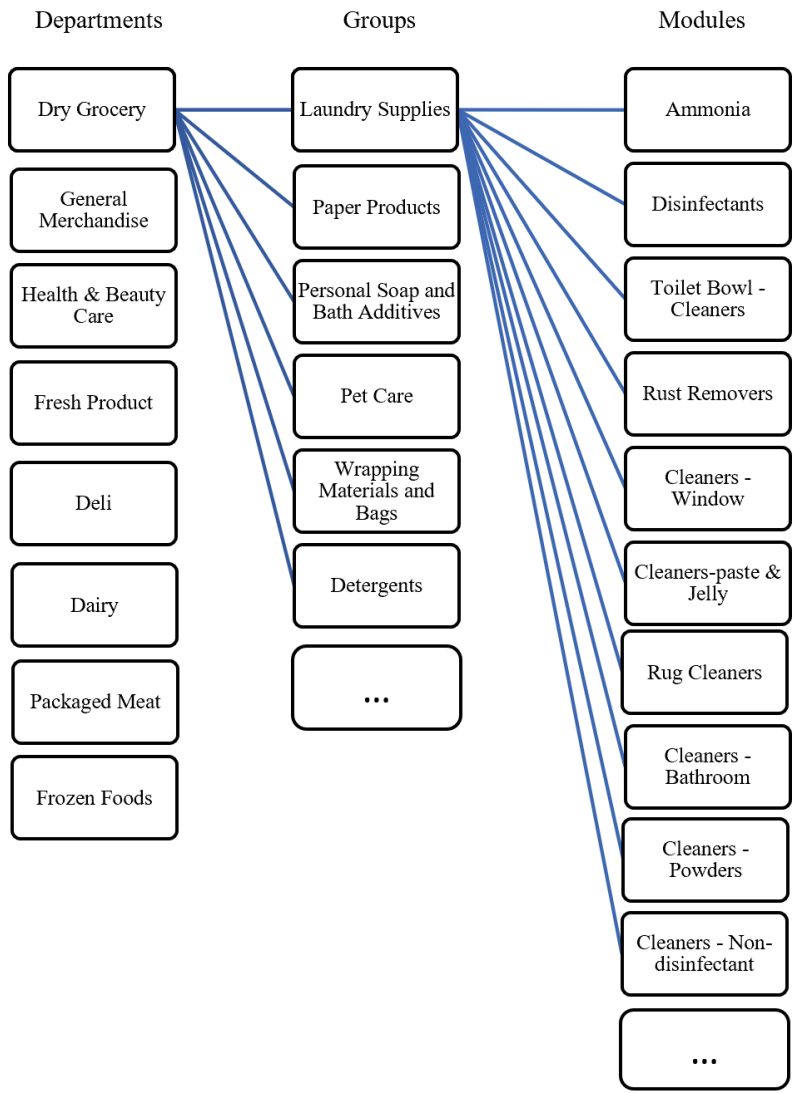

Figure 3. Cross-Category Aggregated Frequency of Price Changes in Cents (NRS Data Full Sample)

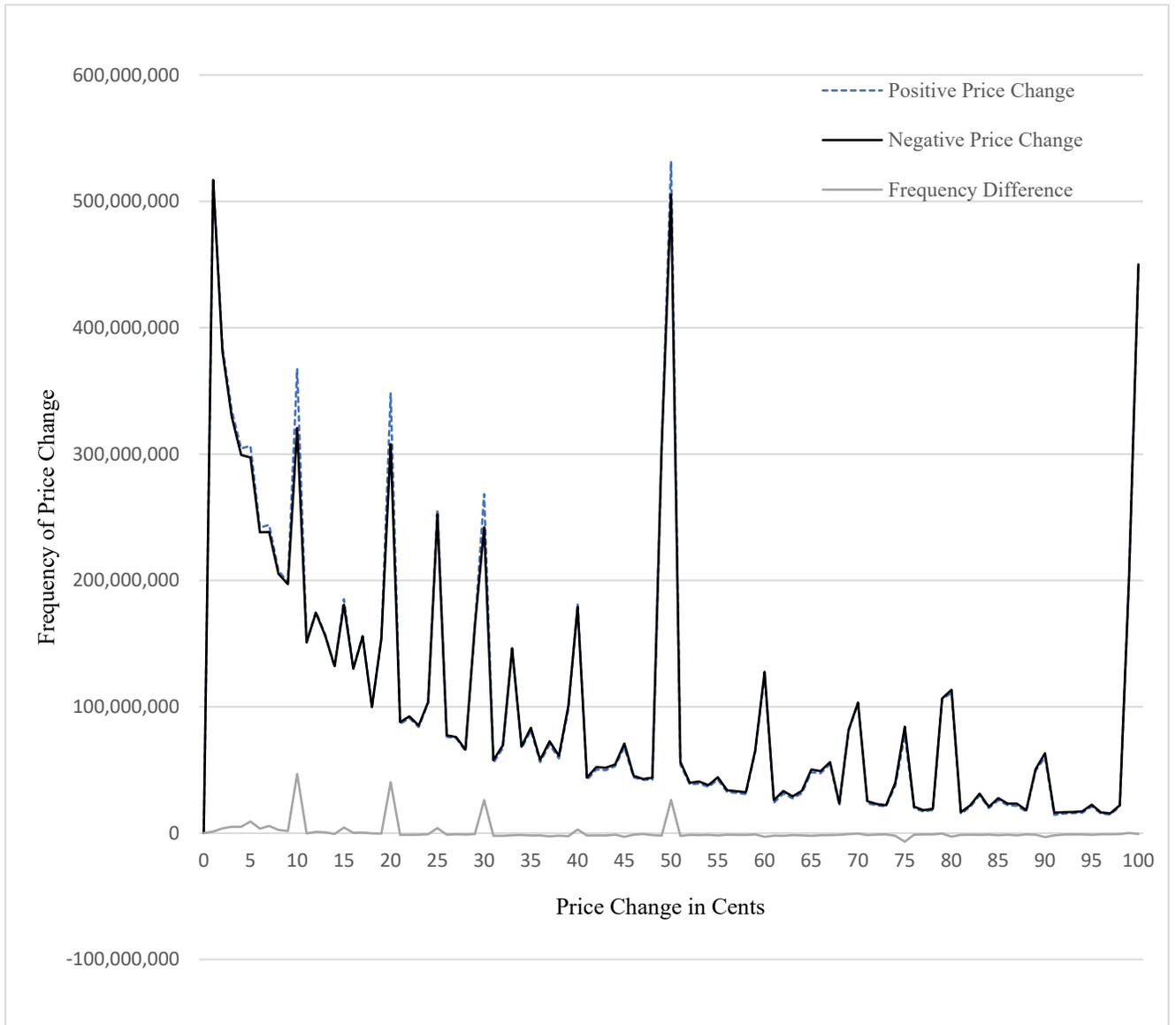

Figure 4. APIS and APIS-R Frequency Distribution at Sub-Category (Module) Level (NRS Data)

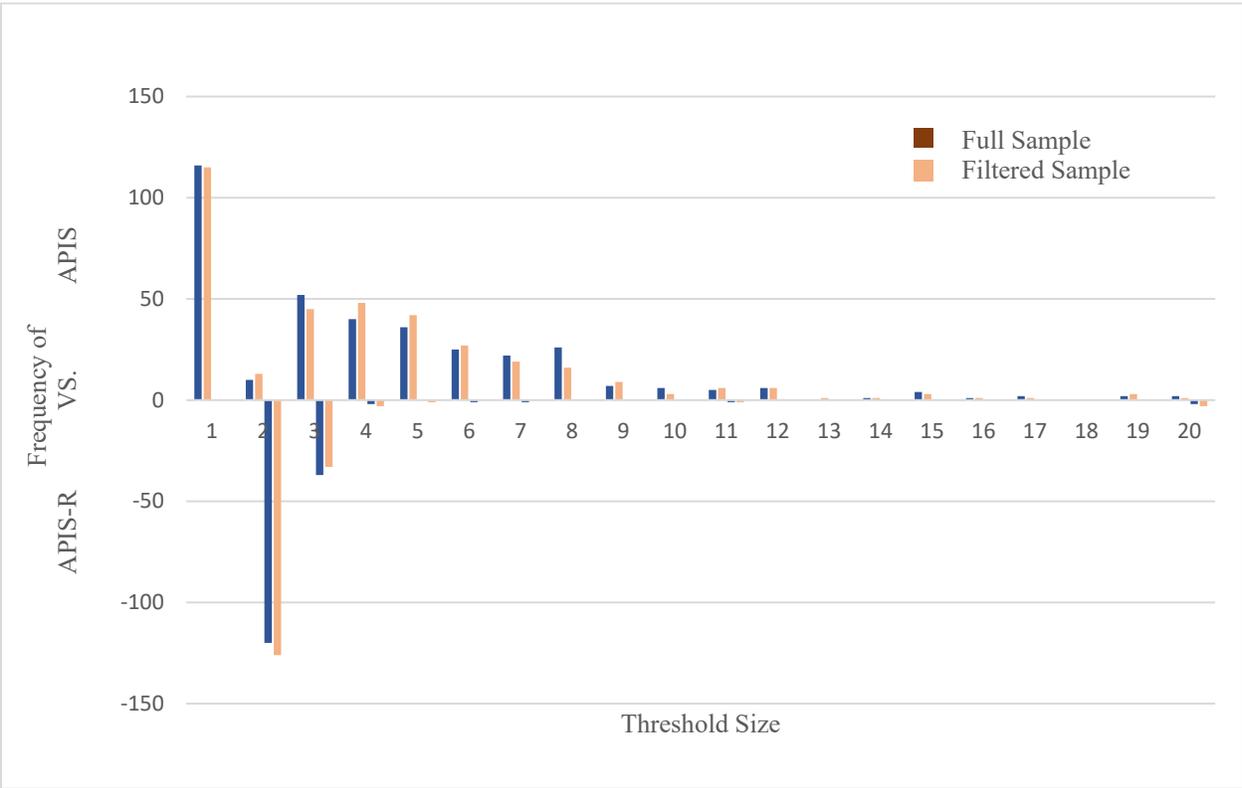

Figure 5. Asymmetry Thresholds Distribution at Retailer Level (NRS Data)

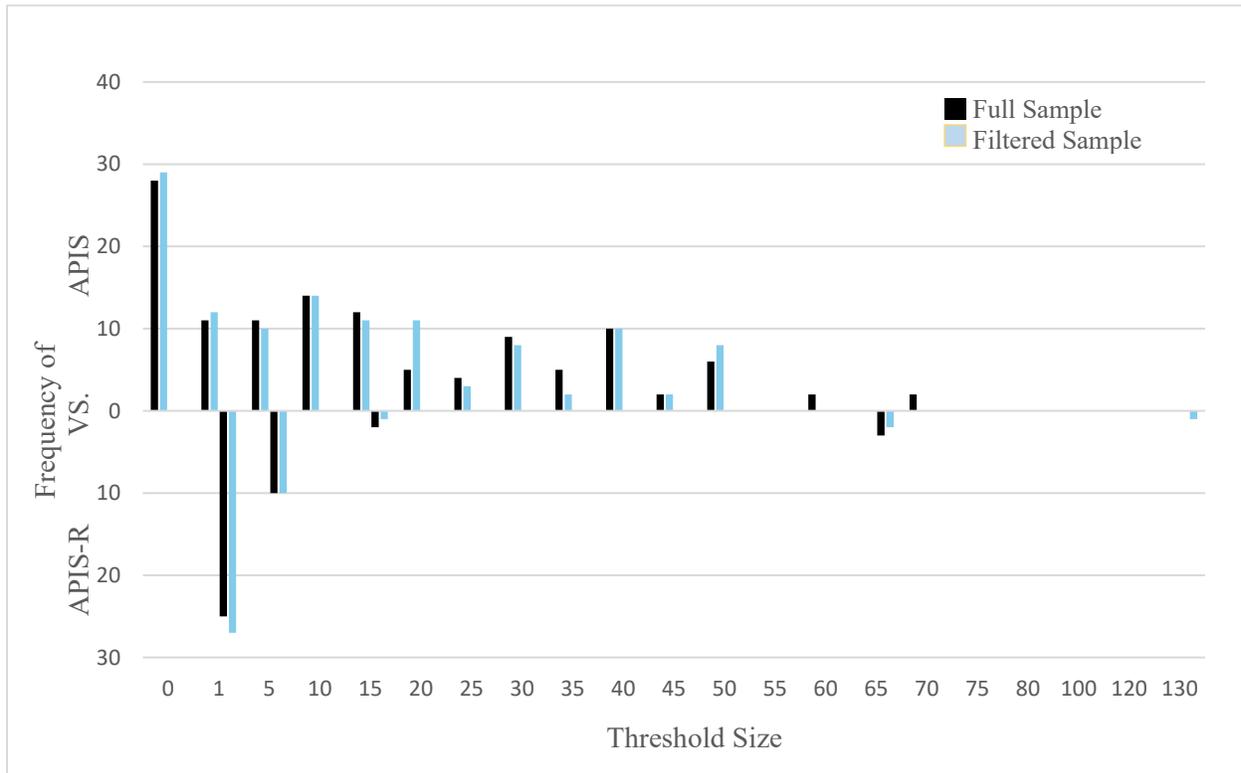

Figure 6. Transaction and Weekly-Aggregated Price Change Frequency Distribution
(Two-Store Transaction Data)

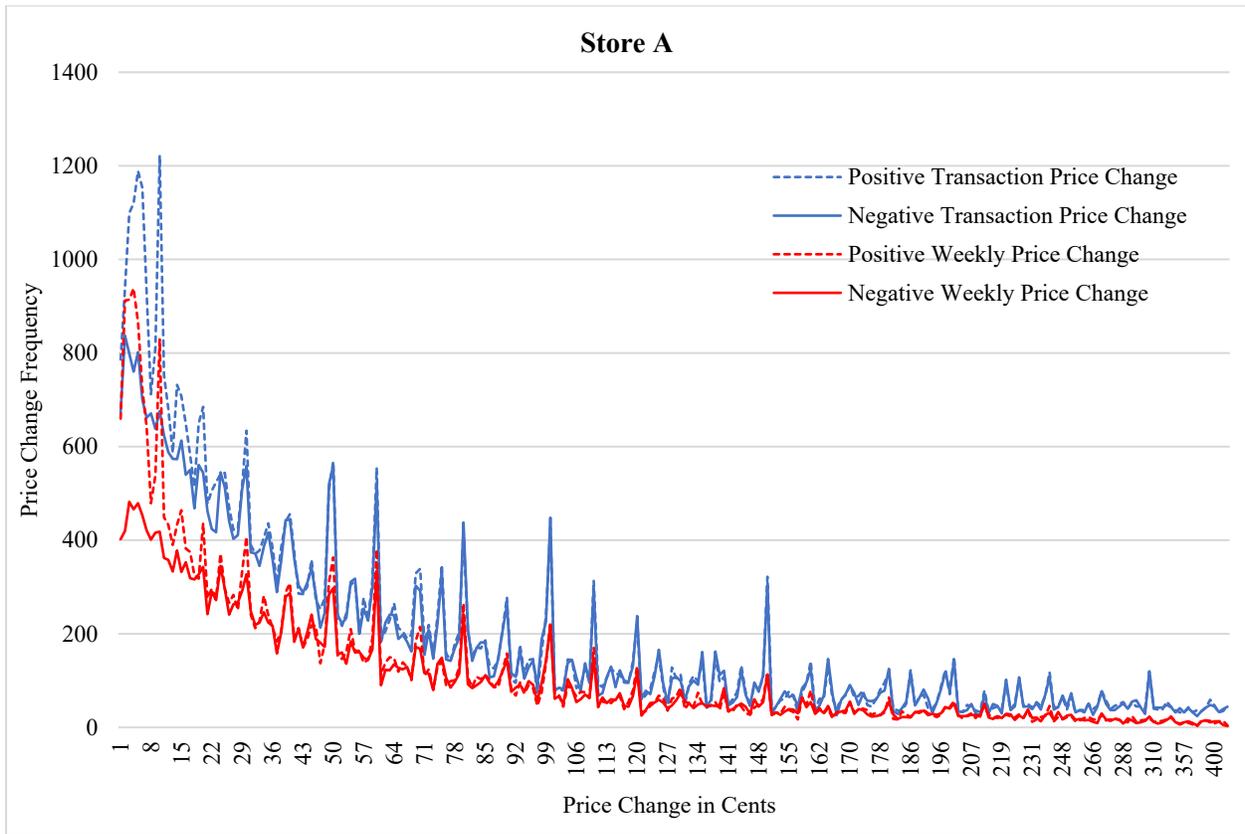

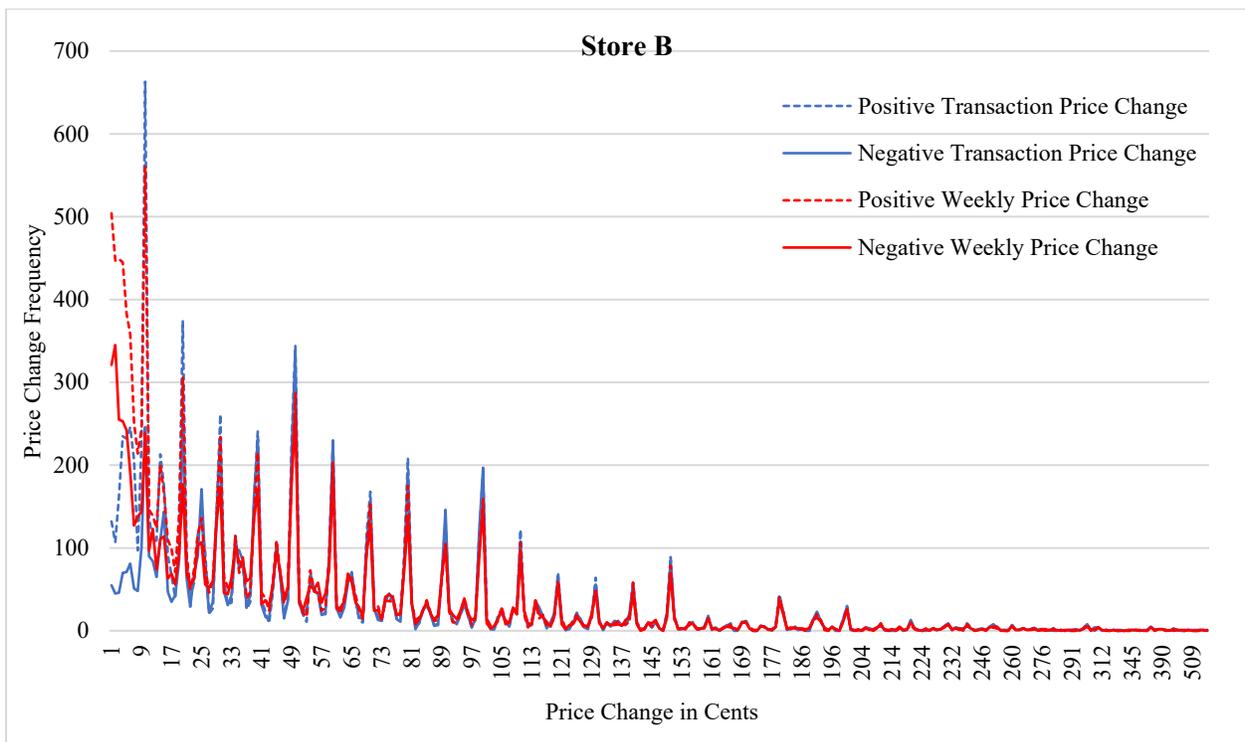

## Table 1. Summary of Analyzed NRS Data at the Product Category (Group) Level

| Department Name | Category (Group) Name | No. of Products | No. of Price Observations | Asy. Threshold (Full Sample) | Asy. Threshold (Filtered Sample) |
|---|---|---|---|---|---|
| Dry Grocery | Candy | 289,747 | 5,907,425,280 | -4 | -9 |
| | Gum | 18,328 | 1,215,013,376 | 0 | 0 |
| | Juice, Drinks - Canned, Bottled | 107,917 | 3,592,290,560 | 20 | 20 |
| | Pet Food | 74,207 | 3,107,025,664 | 14 | 13 |
| | Prepared Food-Ready-To-Serve | 61,235 | 1,550,426,112 | 25 | 25 |
| | Soup | 42,337 | 1,882,395,136 | 30 | 30 |
| | Baking Mixes | 24,596 | 781,248,704 | 9 | 8 |
| | Breakfast Food | 24,107 | 1,125,582,080 | -1 | -1 |
| | Cereal | 38,266 | 1,807,372,672 | 35 | 31 |
| | Coffee | 61,510 | 1,229,685,120 | 0 | -1 |
| | Desserts, Gelatins, Syrup | 20,902 | 965,305,024 | 7 | 7 |
| | Nuts | 65,059 | 1,023,383,232 | -2 | -1 |
| | Packaged Milk and Modifiers | 17,206 | 649,159,424 | 16 | 16 |
| | Sugar, Sweeteners | 9,422 | 298,786,784 | 21 | 22 |
| | Tea | 61,210 | 1,257,548,928 | 10 | 10 |
| | Bread and Baked Goods | 299,239 | 2,605,746,688 | 60 | 39 |
| | Cookies | 102,462 | 1,891,238,912 | -1 | 25 |
| | Crackers | 30,165 | 1,107,950,976 | 41 | 50 |
| | Snacks | 213,708 | 4,488,684,544 | 35 | 14 |
| | Soft Drinks-Non-Carbonated | 59,337 | 1,618,847,360 | -1 | -1 |
| Frozen Foods | Baked Goods-Frozen | 18,804 | 437,997,856 | 21 | 14 |
| | Breakfast Foods-Frozen | 14,016 | 512,820,640 | -1 | -1 |
| | Ice Cream, Novelties | 91,355 | 2,096,284,544 | 30 | 13 |
| | Juices, Drinks-Frozen | 3,441 | 157,345,696 | 24 | 23 |
| | Pizza/Snacks/Hors Doeurves-Frzn | 39,026 | 944,168,256 | 0 | -1 |
| | Prepared Foods-Frozen | 96,910 | 2,701,307,648 | 25 | 23 |
| | Unprep Meat/Poultry/Seafood-Frzn | 39,922 | 325,982,144 | 78 | 77 |
| Dairy | Cheese | 89,513 | 1,844,350,464 | 32 | 14 |
| | Eggs | 9,941 | 94,186,312 | 34 | 34 |
| | Milk | 58,193 | 746,784,320 | 21 | 21 |
| | Snacks, Spreads, Dips-Dairy | 32,946 | 336,883,488 | 14 | 13 |
| | Yogurt | 36,829 | 1,207,518,976 | 5 | 5 |
| Deli | Dressings/Salads/Prep Foods-Deli | 128,385 | 1,779,300,864 | -1 | -1 |
| Packaged Meat | Packaged Meats-Deli | 105,075 | 1,786,796,416 | 72 | 66 |
| | Fresh Meat | 11,147 | 122,056,672 | 48 | 48 |
| Fresh Produce | Fresh Produce | 121,681 | 828,927,296 | 48 | 48 |
| Non_Food Grocery | Detergents | 34,141 | 1,644,012,160 | 1 | 2 |
| | Household Cleaners | 35,886 | 1,246,793,472 | 1 | 0 |
| | Laundry Supplies | 44,919 | 1,156,176,640 | -1 | -1 |
| | Paper Products | 178,806 | 2,523,770,112 | -3 | -3 |
| | Personal Soap and Bath Additives | 89,812 | 1,818,637,312 | -4 | -5 |
| | Pet Care | 143,056 | 1,081,385,344 | -4 | -4 |
| | Wrapping Materials and Bags | 28,871 | 954,846,784 | 26 | 28 |
| General Merchandise | Automotive | 23,392 | 291,624,032 | 27 | 28 |
| | Batteries and Flashlights | 39,673 | 673,595,712 | -5 | -5 |
| | Books and Magazines | 13,579 | 541,083,072 | -20 | -12 |
| | Cookware | 49,007 | 348,034,592 | -9 | -9 |
| | Glassware, Tableware | 261,232 | 925,427,968 | -9 | -9 |
| | Kitchen Gadgets | 264,768 | 1,246,782,592 | -9 | -9 |
| | Toys & Sporting Goods | 22,885 | 25,291,164 | -2 | -2 |
| Health & Beauty Care | Baby Needs | 52,799 | 680,964,096 | -6 | -6 |
| | Hair Care | 182,990 | 3,714,681,344 | -9 | -6 |
| | Medications/Remedies/Health Aids | 1,879 | 89,118,656 | 6 | 6 |
| | Oral Hygiene | 50,977 | 2,499,003,648 | -5 | -4 |
| | Skin Care Preparations | 117,057 | 1,905,587,200 | -4 | -4 |
| | Vitamins | 157,775 | 1,907,149,312 | -3 | -3 |
| Total # | 56 | 4,311,648 | 79,301,793,380 | | |
| Avg. APIS | | | | 27.0 | 24.9 |
| Avg. APIS-R | | | | 4.7 | 4.3 |
| APIS Count | | | | 31 | 31 |
| APIS-R Count | | | | 22 | 23 |

Note: a negative value in the Asymmetry Threshold columns indicates an APIS-R.

Table 2. Asymmetric Price Change Thresholds at the Department Level for Inflation Analyses

| Department Name | Full Sample | Filtered Sample | Low Inflation Sample (PPI) | Deflation Sample (PPI) | Inflation Sample (PPI) |
|---|---|---|---|---|---|
| HEALTH & BEAUTY CARE | -6 | -6 | -2 | -6 | -5 |
| DRY GROCERY | 11 | 10 | 1 | -1 | 13 |
| FROZEN FOODS | 30 | 26 | 11 | 11 | 21 |
| DAIRY | 10 | 8 | 7 | 10 | 10 |
| DELI | -1 | -1 | -9 | -1 | -1 |
| PACKAGED MEAT | 72 | 66 | 21 | 50 | 38 |
| FRESH PRODUCE | 48 | 48 | 68 | 48 | 68 |
| NON_FOOD GROCERY | -3 | -3 | -4 | -3 | 3 |
| GENERAL MERCHANDISE | -9 | -9 | -3 | -9 | -9 |
| Avg. APIS | 34.2 | 31.6 | 21.6 | 29.8 | 25.5 |
| Avg. APIS-R | 4.8 | 4.8 | 4.5 | 4.0 | 5.0 |
| APIS Count | 5 | 5 | 5 | 4 | 6 |
| APIS-R Count | 4 | 4 | 4 | 5 | 3 |

Note: a negative value in the threshold field indicates an APIS-R threshold.

# WEB APPENDIX

Retail Price Ripples

November 30, 2025

*Analysis A1: Excluding 1¢ Price Changes to account for rounding*

In calculating price changes in the NRS data, we had to round up non-integer price changes to the next integer, and thus all smaller-than-1¢ price changes appear as 1¢ changes. This may lead to an inaccurate measurement of some thresholds, because the asymmetry thresholds are identified by the first price-change where asymmetry is not statistically supported, starting from 1¢. We indeed observe a large number of 1¢ price changes. We also find that a large portion of 1¢ APIS-R thresholds, which may be due to this noise – e.g., we find 54 1¢ APIS-R thresholds at the sub-category (module) level, comprising 10.2% of all the sub-categories, and 32.9% of all APIS-R thresholds.

To check the impact of such rounding, we re-do all the estimations in Analysis 1 using the same samples but with all 1¢ price changes excluded. This results in removing about 1.03 billion price changes – about 1.3% of the erstwhile calculations. We find larger average thresholds and a larger portion of APIS thresholds at almost all aggregation levels. At the sub-category level, 58.8% of the thresholds are APIS compared to 46.9% in Analysis 1. The average thresholds are about 18.2¢ and 11¢ for APIS and APIS-R, respectively (see Table A2 in the Web Appendix).

At the product category (group) level, 17 categories exhibit APIS-R and 37 categories (66%) exhibit APIS, compared with 55.4% in Analysis 1. The overall average threshold is about 14.9¢, 1.8¢ higher than in Analysis 1. At the category level, the average APIS and APIS-R thresholds are 25.3¢ and 6.1¢, respectively. At the product department level, all categories remain the same except for Deli, the threshold of which changes from APIS-R 1¢ to APIS 21¢.

To summarize, after excluding 1¢ price changes, we observe more APIS thresholds. Most 1¢ APIS-R thresholds turn into significant APIS thresholds, rather than APIS-R thresholds of other size, or no-asymmetry.  This indicates that 1¢ price changes may contribute to the underestimation of asymmetry, especially APIS.

### Analysis A2. Alternative Measures of Inflation

Since the earlier analyses used PPI inflation rates, we repeat them using inflation rates from CPI.  The findings are similar.  The average APIS threshold in the low inflation samples is 7.7¢ at the sub-category level and 12¢ at the category level.  The average APIS threshold for the deflation period is 13.2¢ at the sub-category level, and 20.1¢ at the category level.  During periods with a larger than 0.5% CPI inflation rates, the average category level threshold is still close to the results we obtained using the PPI inflation (see Table A2 for category level results).

### Analysis A3. Lagged Price Adjustment

We allow for lagged price adjustment and repeat the analysis with 4-, 8-, 12- and 16- week lags after the PPI-deflationary periods.  The results show that the asymmetry still holds at different aggregation levels, averaging an APIS threshold of 9.2¢ at the sub-category level and 12.5¢ at the category level when 4-week lag is applied.  (See Table A3).

### Analysis A4. First Year Sample versus Last Year Sample

Finally, to further control for the effects of inflation, we compare the asymmetry thresholds during the first year of our sample period with those during the last year, at the sub-category level.  Since there is an upward inflation trend during 2006–2015, we are supposed to see stronger APIS in the last year of the sample period if inflation is causing the asymmetry.  The results indicate that in 247 of the 512 sub-categories (the number of sub-categories varies each year), an APIS threshold is found in the first-year sample, but only 146 APIS thresholds (out of 524 sub-categories) are identified in the last year. Most APIS thresholds are smaller in the last year compared with the first year (an average of 10.5¢ in the first year vs. 4.4¢ in the last year). We get similar results at the retailer level. Among the 93 retailers in the first year, for 56 (60.2%) of them we find APIS. That number drops to 42 (40.4% out of 104 retailers) in the last year. The average APIS threshold also decreases from 12.5¢ in the first year to 9.7¢ in the last year at retailer level.

Table A1. Asymmetric Price Change Thresholds Sub-Category (Module) Level Summary

| Summary | Full Sample | Filtered Sample | Low Inflation Sample (PPI) | Deflation Sample (PPI) | Inflation Sample (PPI) |
|---|---|---|---|---|---|
| N | 527 | 527 | 521 | 527 | 527 |
| Avg. APIS Threshold | 18.1 | 17.1 | 5.3 | 11.8 | 12.6 |
| Avg. APIS-R Threshold | 7.4 | 18.6 | 3.8 | 4.4 | 5.3 |
| APIS Count | 247 | 245 | 146 | 218 | 275 |
| APIS-R Count | 164 | 167 | 141 | 176 | 142 |

Note: More details about module-level thresholds are available from the authors upon request.

Table A2. Asymmetric Thresholds at Product Category Level for Alternative Inflation Analyses

| Department Name | Category Name | Full Sample | Filtered Sample | PPI Low-Inflation | PPI Deflation | CPI Low-Inflation | CPI Deflation |
|---|---|---|---|---|---|---|---|
| Dry Grocery | Candy | -4 | -9 | 1 | -3 | -3 | -3 |
| | Gum | 0 | 0 | 1 | -2 | 1 | 11 |
| | Juice, Drinks - Canned, Bottled | 20 | 20 | -8 | 30 | 10 | 0 |
| | Pet Food | 14 | 13 | 6 | -3 | -3 | -3 |
| | Prepared Food-Ready-To-Serve | 25 | 25 | -3 | 17 | -1 | 34 |
| | Soup | 30 | 30 | 14 | 0 | 6 | 0 |
| | Baking Mixes | 9 | 8 | -3 | -3 | 0 | -3 |
| | Breakfast Food | -1 | -1 | 2 | -1 | -2 | 13 |
| | Cereal | 35 | 31 | 2 | -1 | 7 | -1 |
| | Coffee | 0 | -1 | 23 | -1 | -1 | -1 |
| | Desserts, Gelatins, Syrup | 7 | 7 | 1 | 1 | 1 | 2 |
| | Nuts | -2 | -1 | -2 | -2 | -2 | 0 |
| | Packaged Milk and Modifiers | 16 | 16 | 11 | -1 | 0 | -9 |
| | Sugar, Sweeteners | 21 | 22 | -1 | 8 | 29 | 0 |
| | Tea | 10 | 10 | 3 | 0 | 9 | 9 |
| | Bread and Baked Goods | 60 | 39 | -19 | 29 | 10 | 32 |
| | Cookies | -1 | 25 | -2 | -1 | -1 | -1 |
| | Crackers | 41 | 50 | -1 | 13 | 0 | 9 |
| | Snacks | 35 | 14 | -8 | 10 | 11 | 10 |
| | Soft Drinks-Non-Carbonated | -1 | -1 | 4 | -1 | 5 | -1 |
| Frozen Foods | Baked Goods-Frozen | 21 | 14 | 2 | 11 | 14 | 11 |
| | Breakfast Foods-Frozen | -1 | -1 | 1 | -1 | -2 | -1 |
| | Ice Cream, Novelties | 30 | 13 | 1 | 15 | 5 | 15 |
| | Juices, Drinks-Frozen | 24 | 23 | 0 | 0 | 2 | 15 |
| | Pizza/Snacks/Hors Doeurves-Frzn | 0 | -1 | 9 | 6 | 3 | -1 |
| | Prepared Foods-Frozen | 25 | 23 | 29 | 3 | 10 | 8 |
| | Unprep Meat/Poultry/Seafood-Frzn | 78 | 77 | -2 | 58 | 18 | 59 |
| Dairy | Cheese | 32 | 14 | 2 | 13 | 9 | 26 |
| | Eggs | 34 | 34 | 0 | 79 | 35 | 39 |
| | Milk | 21 | 21 | 12 | 16 | 21 | 26 |
| | Snacks, Spreads, Dips-Dairy | 14 | 13 | 14 | -1 | -1 | -1 |
| | Yogurt | 5 | 5 | 6 | 5 | 5 | 5 |
| Deli | Dressings/Salads/Prep Foods-Deli | -1 | -1 | -9 | -1 | -2 | 0 |
| Packaged Meat | Packaged Meats-Deli | 72 | 66 | 21 | 50 | 39 | 50 |
| | Fresh Meat | 48 | 48 | 0 | 48 | 39 | 48 |
| Fresh Produce | Fresh Produce | 48 | 48 | 68 | 48 | 28 | 48 |
| Non_Food Grocery | Detergents | 1 | 2 | -4 | -4 | -1 | -4 |
| | Household Cleaners | 1 | 0 | -1 | -2 | 10 | 0 |
| | Laundry Supplies | -1 | -1 | 2 | 1 | -1 | -4 |
| | Paper Products | -3 | -3 | -4 | -3 | 1 | -3 |
| | Personal Soap and Bath Additives | -4 | -5 | -5 | -5 | -1 | -5 |
| | Pet Care | -4 | -4 | -1 | -4 | -1 | 0 |
| | Wrapping Materials and Bags | 26 | 28 | -3 | 4 | 5 | 3 |
| General Merchandise | Automotive | 27 | 28 | 18 | -1 | 0 | -1 |
| | Batteries and Flashlights | -5 | -5 | -2 | -3 | -7 | 1 |
| | Books and Magazines | -20 | -12 | 0 | -12 | -7 | -37 |
| | Cookware | -9 | -9 | 0 | -4 | 1 | -1 |
| | Glassware, Tableware | -9 | -9 | 1 | 0 | -9 | 0 |
| | Kitchen Gadgets | -9 | -9 | 1 | -9 | -9 | -9 |
| | Toys & Sporting Goods | -2 | -2 | 0 | 0 | 0 | 2 |
| Health & Beauty Care | Baby Needs | -6 | -6 | -1 | 1 | -1 | -4 |
| | Hair Care | -9 | -6 | -2 | 1 | -4 | -3 |
| | Medications/Remedies/Health Aids | 6 | 6 | 0 | 6 | 0 | 6 |
| | Oral Hygiene | -5 | -4 | -8 | -5 | 2 | -5 |
| | Skin Care Preparations | -4 | -4 | 2 | -9 | -1 | -9 |
| | Vitamins | -3 | -3 | -2 | -3 | -5 | -3 |
| Avg. APIS | | 27.0 | 24.9 | 9.5 | 19.7 | 12.0 | 20.1 |
| Avg. APIS-R | | 4.7 | 4.3 | 4.1 | 3.2 | 3.0 | 4.7 |
| APIS Count | | 31 | 31 | 27 | 24 | 28 | 24 |
| APIS-R Count | | 22 | 23 | 22 | 27 | 22 | 24 |

Note: A negative value in the Asymmetry Threshold columns indicates an APIS-R.

Table A3. Asymmetric Thresholds at Product Category Level for Lagged Price Changes (PPI)

| Category Name | PPI Low-Inflation | PPI Deflation | 4-week Lag Deflation Period | 8-week Lag Deflation Period | 12-week Lag Deflation Period | 16-week Lag Deflation Period |
|---|---|---|---|---|---|---|
| Candy | 1 | -3 | -9 | -9 | -9 | -9 |
| Gum | 1 | -2 | -4 | 1 | 1 | 0 |
| Juice, Drinks - Canned, Bottled | -8 | 30 | 20 | 0 | 0 | 20 |
| Pet Food | 6 | -3 | 8 | -5 | -3 | 1 |
| Prepared Food-Ready-To-Serve | -3 | 17 | 16 | 17 | -1 | 0 |
| Soup | 14 | 0 | 19 | 33 | 6 | 30 |
| Baking Mixes | -3 | -3 | 5 | 0 | -4 | 5 |
| Breakfast Food | 2 | -1 | -1 | -1 | -1 | -1 |
| Cereal | 2 | -1 | 11 | 3 | 6 | 0 |
| Coffee | 23 | -1 | 0 | -1 | -1 | -1 |
| Desserts, Gelatins, Syrup | 1 | 1 | 6 | 7 | 2 | 7 |
| Nuts | -2 | -2 | -3 | 8 | 3 | 3 |
| Packaged Milk and Modifiers | 11 | -1 | 10 | -1 | -1 | 11 |
| Sugar, Sweeteners | -1 | 8 | 17 | 16 | 8 | 10 |
| Tea | 3 | 0 | -1 | -1 | -1 | 11 |
| Bread and Baked Goods | -19 | 29 | 14 | 18 | 19 | 32 |
| Cookies | -2 | -1 | -1 | -1 | -1 | -1 |
| Crackers | -1 | 13 | 2 | 6 | 5 | -1 |
| Snacks | -8 | 10 | 13 | 8 | 10 | 14 |
| Soft Drinks-Non-Carbonated | 4 | -1 | -1 | -2 | -1 | -1 |
| Baked Goods-Frozen | 2 | 11 | 11 | 13 | 12 | 11 |
| Breakfast Foods-Frozen | 1 | -1 | -1 | -1 | -2 | -1 |
| Ice Cream, Novelties | 1 | 15 | 0 | -1 | -1 | 9 |
| Juices, Drinks-Frozen | 0 | 0 | -1 | 22 | 19 | 32 |
| Pizza/Snacks/Hors Doeurves-Frzn | 9 | 6 | 8 | -1 | -1 | 13 |
| Prepared Foods-Frozen | 29 | 3 | 7 | -1 | -1 | 9 |
| Unprep Meat/Poultry/Seafood-Frzn | -2 | 58 | 13 | 50 | 2 | 32 |
| Cheese | 2 | 13 | 5 | 13 | 28 | 15 |
| Eggs | 0 | 79 | 31 | 48 | 48 | 49 |
| Milk | 12 | 16 | 10 | 12 | 24 | 21 |
| Snacks, Spreads, Dips-Dairy | 14 | -1 | 0 | 0 | -1 | 11 |
| Yogurt | 6 | 5 | 6 | 5 | 5 | 5 |
| Dressings/Salads/Prep Foods-Deli | -9 | -1 | -1 | -1 | -1 | -1 |
| Packaged Meats-Deli | 21 | 50 | 32 | 51 | 41 | 65 |
| Fresh Meat | 0 | 48 | 48 | 49 | 48 | 48 |
| Fresh Produce | 68 | 48 | 29 | 72 | 60 | 48 |
| Detergents | -4 | -4 | 3 | 2 | -4 | 1 |
| Household Cleaners | -1 | -2 | -2 | 2 | 10 | 5 |
| Laundry Supplies | 2 | 1 | 11 | -2 | -4 | 6 |
| Paper Products | -4 | -3 | -5 | -9 | -4 | -1 |
| Personal Soap and Bath Additives | -5 | -5 | -4 | 1 | 1 | -7 |
| Pet Care | -1 | -4 | 0 | -4 | -3 | 1 |
| Wrapping Materials and Bags | -3 | 4 | 2 | 8 | -1 | 10 |
| Automotive | 18 | -1 | 11 | 21 | 0 | 21 |
| Batteries and Flashlights | -2 | -3 | -2 | 1 | -9 | -9 |
| Books and Magazines | 0 | -12 | 0 | -12 | -10 | -9 |
| Cookware | 0 | -4 | -20 | -14 | -9 | -9 |
| Glassware, Tableware | 1 | 0 | 1 | -9 | -9 | 0 |
| Kitchen Gadgets | 1 | -9 | -9 | -9 | -9 | -9 |
| Toys & Sporting Goods | 0 | 0 | -2 | -1 | 2 | 2 |
| Baby Needs | -1 | 1 | 1 | -2 | -6 | 1 |
| Hair Care | -2 | 1 | -3 | -5 | 1 | -9 |
| Medications/Remedies/Health Aids | 0 | 6 | 6 | 7 | 4 | 8 |
| Oral Hygiene | -8 | -5 | -6 | -3 | -7 | -3 |
| Skin Care Preparations | 2 | -9 | -3 | -3 | -1 | -3 |
| Vitamins | -2 | -3 | -3 | -1 | -3 | -3 |
| Avg. APIS | 9.5 | 19.7 | 12.5 | 18.3 | 15.2 | 16.7 |
| Avg. APIS-R | 4.1 | 3.2 | 3.9 | 3.8 | 3.6 | 4.3 |
| APIS Count | 27 | 24 | 30 | 27 | 24 | 34 |
| APIS-R Count | 22 | 27 | 21 | 26 | 30 | 18 |

Note: A negative value in the Asymmetry Threshold columns indicates an APIS-R.

Table A4. Asymmetric Price Change Thresholds at Dept. Level for the Transaction Price Data

| Dept. Name | Store A | | Store B | |
| --- | --- | --- | --- | --- |
| | Transaction Price | Weekly Aggregate Price | Transaction Price | Weekly Aggregate Price |
| Canned food | 0 | 11 | 11 | 11 |
| Canned products (not food) | 0 | 6 | 0 | 1 |
| Butcher's shop | 0 | 0 | 0 | 2 |
| Fruit and vegetables | 0 | 0 | 0 | 0 |
| Delicatessen | 0 | 0 | 0 | 0 |
| Dairy products | 1 | 2 | 0 | 0 |
| Bread | 0 | 0 | | |
| Deep-frozen food | 0 | 0 | | |
| Fishmonger's | 0 | 0 | | |
| General store/ steward's office | 0 | 0 | | |
| Affiliation | 0 | 0 | | |
| Low level pharmacy/ Newspapers | 0 | 0 | | |
| General store | 0 | 0 | | |
| Textile/ household linen | 0 | 0 | | |
| Housewares | 0 | 0 | | |
| Toys | 0 | 0 | | |
| Stationery store | 0 | 0 | | |
| Underwear | 7 | 7 | | |
| Support department | 0 | 0 | | |
| **Avg.** | **0.4** | **1.4** | **1.8** | **2.3** |

Table A5. Asymmetric Price Change Thresholds at Dept-Year Level for the Transaction Price Data

| Dept. Name | Year | Store A | | Store B | |
| --- | --- | --- | --- | --- | --- |
| | | Transaction Price | Weekly Aggregate Price | Transaction Price | Weekly Aggregate Price |
| Canned food | 2007 | 0 | 7 | 0 | 0 |
| Canned products (not food) | 2007 | 0 | 6 | 0 | 0 |
| Butcher's shop | 2007 | 0 | 0 | 3 | 1 |
| Fruit and vegetables | 2007 | 0 | 0 | 0 | 0 |
| Delicatessen | 2007 | 0 | 0 | 0 | 1 |
| Dairy products | 2007 | 3 | 1 | 0 | 0 |
| Bread | 2007 | 0 | 0 | | |
| Deep-frozen food | 2007 | 0 | 0 | | |
| Fishmonger's | 2007 | 0 | 0 | | |
| General store/ steward's office | 2007 | 0 | 0 | | |
| Low level pharmacy/ Newspapers | 2007 | 0 | 0 | | |
| General store | 2007 | 0 | 0 | | |
| Textile/ household linen | 2007 | 0 | 0 | | |
| Housewares | 2007 | 0 | 0 | | |
| Stationery store | 2007 | 10 | 10 | | |
| Underwear | 2007 | 0 | 0 | | |
| Support department | 2007 | 0 | 0 | | |
| Canned food | 2008 | 17 | 11 | 11 | 21 |
| Canned products (not food) | 2008 | 0 | 0 | 0 | 0 |
| Butcher's shop | 2008 | 0 | 0 | 0 | 2 |
| Fruit and vegetables | 2008 | 1 | 0 | 0 | 0 |
| Delicatessen | 2008 | 0 | 0 | 0 | -1 |
| Dairy products | 2008 | 0 | 0 | 0 | 0 |
| Bread | 2008 | 0 | 0 | | |
| Deep-frozen food | 2008 | 0 | 0 | | |
| Fishmonger's | 2008 | 0 | 0 | | |
| General store/ steward's office | 2008 | 0 | 0 | | |
| Affiliation | 2008 | 0 | 0 | | |
| General store | 2008 | 0 | 0 | | |
| Textile/ household linen | 2008 | 0 | 0 | | |
| Housewares | 2008 | 0 | 0 | | |
| Toys | 2008 | 0 | 0 | | |
| Stationery store | 2008 | 0 | 0 | | |
| Underwear | 2008 | 7 | 7 | | |
| Support department | 2008 | 0 | 0 | | |
| **Avg.** | | **1.1** | **1.2** | **1.2** | **2.0** |

Table A6. Asymmetric Price Change Thresholds at Yearly Level (NRS Data)

| Year | Threshold | CPI Inflation Rate (%) | PPI Inflation Rate (%) |
|------|-----------|------------------------|------------------------|
| 2006 | 21 | 3.23 | 1.6 |
| 2007 | 21 | 2.85 | 7.85 |
| 2008 | 37 | 3.84 | -4.31 |
| 2009 | -4 | -0.36 | 4.21 |
| 2010 | -1 | 1.64 | 6.51 |
| 2011 | 30 | 3.16 | 5.32 |
| 2012 | -1 | 2.07 | 0.85 |
| 2013 | -9 | 1.46 | 0.25 |
| 2014 | 8 | 1.62 | -2.48 |
| 2015 | -2 | 0.12 | -6.85 |
| **Avg.** | **10.0** | **2.0** | **1.3** |

Note: A negative value in the threshold field indicates an APIS-R threshold.

Table A7. Comparison of Asymmetry Threshold Results with Chen et al. (2008)

| Product Categories | | Full Sample | | Low Inflation Sample | | Deflation Sample | |
|---|---|---|---|---|---|---|---|
| Chen et al. 2008 | This paper | Chen et al. 2008 | This paper | Chen et al. 2008 | This paper | Chen et al. 2008 | This paper |
| Analgesics | Tooth & Gum Analgesics (sub-category) | 30 | 6 | 10 | 0 | 10 | 6 |
| Bath soap | Soap- Bar (sub-category) | 6 | -3 | 0 | -8 | 0 | -4 |
| Bathroom tissues | TOILET TISSUE (sub-category) | 6 | 25 | 4 | 0 | 4 | 3 |
| Bottled juices | JUICE, DRINKS - CANNED, BOTTLED (category) | 12 | 20 | 15 | -8 | 12 | 30 |
| Canned soup | Soup- Canned (sub-category) | 12 | 30 | 12 | 9 | 10 | 11 |
| Canned tuna | | 1 | | 2 | | 1 | |
| Cereals | Cereal (category) | 29 | 35 | 24 | 2 | 1 | -1 |
| Cheeses | Cheese (category) | 9 | 32 | 9 | 2 | 9 | 12 |
| Cookies | Cookies (category) | 11 | -1 | 11 | -2 | 9 | -1 |
| Crackers | Crackers (category) | 10 | 41 | 2 | -1 | 4 | 13 |
| Dish detergent | Automatic Dishwasher Compounds (sub-category) | 5 | -2 | 4 | 0 | 6 | -2 |
| Fabric softeners | FABRIC SOFTENERS- LIQUID (sub-category) | 5 | 0 | 11 | 3 | 7 | 0 |
| Front-end-candies | CANDY (category) | 5 | -4 | 5 | 1 | 5 | -3 |
| Frozen dinners | Dinners-Frozen (sub-category) | 2 | -3 | 10 | 0 | 6 | 3 |
| Frozen entrees | ENTREES - MEAT - 1 FOOD - FROZEN (sub-category) | 20 | 8 | 22 | 11 | 0 | 2 |
| Frozen juices | Juices, Drinks-Frozen (category) | 9 | 24 | 9 | 0 | 10 | 0 |
| Grooming products | HAIR CARE (category) | 20 | -9 | 12 | -2 | 12 | 1 |
| Laundry detergents | Detergents (category) | 16 | 1 | 13 | -4 | 17 | -4 |
| Oatmeal | | 25 | | 2 | | 5 | |
| Paper towels | Paper Towels (sub-category) | 2 | 11 | 2 | 1 | 2 | -1 |
| Refrigerated juices | | 15 | | 9 | | 6 | |
| Shampoos | SHAMPOO-AEROSOL/ LIQUID/ LOTION/ POWDER (sub-category) | 0 | -9 | 10 | -1 | 10 | 1 |
| Snack crackers | Crackers - Flavored Snack (sub-category) | 11 | 0 | 2 | -2 | 2 | 2 |
| Soaps | SOAP - BAR (sub-category) | 1 | -3 | 1 | -8 | 1 | -4 |
| Soft drinks | Soft Drinks-Non-Carbonated (category) | 5 | -1 | 3 | 4 | 5 | -1 |
| Tooth brushes | | 20 | | 3 | | 3 | |
| Tooth pastes | ORAL HYGIENE (category) | 18 | -5 | 14 | -8 | 6 | -5 |
| | **Avg. (all categories)** | **11.3** | **9.0** | **8.2** | **-0.1** | **6.2** | **2.9** |
| | **Avg. (matching categories)** | **11.3** | **9.5** | **8.3** | **0.1** | **5.5** | **1.3** |
| | **Avg. APIS** | 11.7 | **21.2** | 8.5 | **4.1** | 6.5 | **7.6** |
| | **Avg. APIS-R** | N/A | **-4.0** | N/A | **-4.4** | N/A | **-2.6** |